\documentclass[superscriptaddress,amsmath,amssymb,aps,prb,twocolumn,nofootinbib,nolongbibliography]{revtex4-2}
\usepackage{graphicx}
\usepackage{graphics}
\usepackage{dcolumn}          
\usepackage{bm}               
\usepackage{upgreek}
\usepackage{booktabs} 				
\usepackage{tabularx}
\usepackage{array}
\usepackage[colorlinks=true,urlcolor=blue,breaklinks=true]{hyperref}

\usepackage{xcolor}

\usepackage{orcidlink}

\usepackage{xr}

\usepackage{siunitx}
\usepackage{stackengine}
\usepackage{svg}
\usepackage{graphicx,subfigure}
\usepackage{placeins}
\usepackage{bibunits}

\definecolor{green2}{RGB}{15, 117, 19}

\usepackage{titlesec}

\begin{document}

\title{Superconducting Diode Effect in Selectively-Grown Topological Insulator based Josephson Junctions}

\author{Gerrit Behner\,\orcidlink{0000-0002-7218-3841}}
\email{g.behner@fz-juelich.de}
\affiliation{Peter Gr\"unberg Institut (PGI-9), Forschungszentrum J\"ulich, 52425 J\"ulich, Germany}
\affiliation{JARA-Fundamentals of Future Information Technology, J\"ulich-Aachen Research Alliance, Forschungszentrum J\"ulich and RWTH Aachen University, Germany}

\author{Abdur Rehman Jalil
\,\orcidlink{0000-0003-1869-2466}}
\affiliation{Peter Gr\"unberg Institut (PGI-9), Forschungszentrum J\"ulich, 52425 J\"ulich, Germany}
\affiliation{JARA-Fundamentals of Future Information Technology, J\"ulich-Aachen Research Alliance, Forschungszentrum J\"ulich and RWTH Aachen University, Germany}

\author{Detlev Gr\"utzmacher\,\orcidlink{0000-0001-6290-9672}}
\affiliation{Peter Gr\"unberg Institut (PGI-9), Forschungszentrum J\"ulich, 52425 J\"ulich, Germany}
\affiliation{JARA-Fundamentals of Future Information Technology, J\"ulich-Aachen Research Alliance, Forschungszentrum J\"ulich and RWTH Aachen University, Germany}

\author{Thomas Sch\"apers\,\orcidlink{0000-0001-7861-5003}}
\email{th.schaepers@fz-juelich.de}
\affiliation{Peter Gr\"unberg Institut (PGI-9), Forschungszentrum J\"ulich, 52425 J\"ulich, Germany}
\affiliation{JARA-Fundamentals of Future Information Technology, J\"ulich-Aachen Research Alliance, Forschungszentrum J\"ulich and RWTH Aachen University, Germany}

\hyphenation{}
\date{\today}
\begin{abstract}
\noindent 
The Josephson diode effect, where the critical current magnitude depends on its direction, arises when both time-reversal and inversion symmetries are broken - often achieved by a combination of spin-orbit interaction and applied magnetic fields. Taking advantage of the strong spin-orbit coupling inherent in three-dimensional topological insulators, we study this phenomenon in Nb/$\mathrm{Bi_{0.8}Sb_{1.2}Te_3}$/Nb Josephson weak-link junctions. Under an in-plane magnetic field perpendicular to the current direction, we observe a pronounced Josephson diode effect with efficiencies up to 7\%. A crucial component of this behavior is the non-sinusoidal current-phase relationship and an anomalous phase shift, which we attribute to the presence of a ballistic supercurrent component due to the surface states. These findings open up new avenues for harnessing and controlling the Josephson diode effect in topological material systems.
\end{abstract}

\maketitle

\section{Introduction}
\begin{bibunit}[apsrev4-1]
Three-dimensional topological insulators (3D-TIs) are a class of materials that have recently attracted a lot of interest due to its promising applicability in the field of topological quantum computing \cite{Hasan_2010,Ando_2013,Breunig_2021}. The material class exhibits strong spin-orbit coupling. This in turn leads to a band inversion in the bulk electronic band structure. As a consequence, gapless surface states appear, which are protected by time-reversal symmetry. Proximizing a topological insulator nanoribbon with an s-type superconductor and aligning a magnetic field along the nanoribbon gives rise to Majorana zero modes \cite{Lutchyn_2018}. Braiding of these Majorana zero modes is the essential computation operation in topological quantum computing \cite{Nayak_2008,Alicea_2012,Hyart_2013,Sarma_2015,Aasen_2016}. The detection of Majorana zero-modes has proven to be quite difficult, as many initially proposed signatures, like the zero-bias peak and the missing odd Shapiro steps, can in many cases be explained by trivial phenomena. Therefore, new approaches for the detection of Majorana zero-modes are necessary. A key role could be played by the superconducting diode effect which has recently received a lot of attention \cite{Nadeem_2023}. A characteristic of the diode effect is that the magnitude of the critical supercurrent depends on the direction in which the current is driven. The diode effect occurs when both inversion and time-reversal symmetry are broken. A lot of work has been published in semiconductor-superconductor material platforms \cite{Baumgartner_2022,Turini_2022,Costa_2023,Lotfizadeh_2024,Pal2022,Lu2023} where this can be accomplished by the presence of spin-orbit coupling in conjunction with an external magnetic field for the time-reversal symmetry breaking \cite{Reynoso2012,Yokoyama2013,Yokoyama2014,Dolcini2015,Yuan2022,Liu2024,Meyer2024}. Alternatively, the inversion symmetry can be broken by the device layout itself. This can be achieved, for example, by a superconducting quantum interference device (SQUID), where each of the two junctions of the interferometer has a different current-phase relation \cite{Souto_2022,Nikodem2024}. More recently, the asymmetry in a multi-terminal Josephson junction has led to a diode effect, either by keeping one of the superconducting electrodes floating \cite{Gupta_2023,Behner2024}, or by phase biasing using superconducting loops connecting pairs of electrodes in the junction \cite{Coraiola_2024}. For topological insulators specifically, in the last two years many works have proposed unusual behavior of the Josephson diode effect. Here, it is proposed that the magnitude and sign of the diode efficiency is strongly influenced by the external magnetic field, gate voltage and junction length, with the unique properties arising from the interplay of different current-phase relationships across multiple transverse transport channels \cite{Lu2023}. Most importantly, the diode effect in a 3D-TI Josephson junction is proposed as a detector mechanism for the presence of 4$\pi$-periodic contribution of Majorana bound states. Their presence leads to significant enhancement of the superconducting diode effect when the device enters the topological phase and could therefore be used as an indicator of the phase transition \cite{Legg2023}.

\begin{figure*}[t!]
    \centering
    \includegraphics[width=0.99\linewidth]{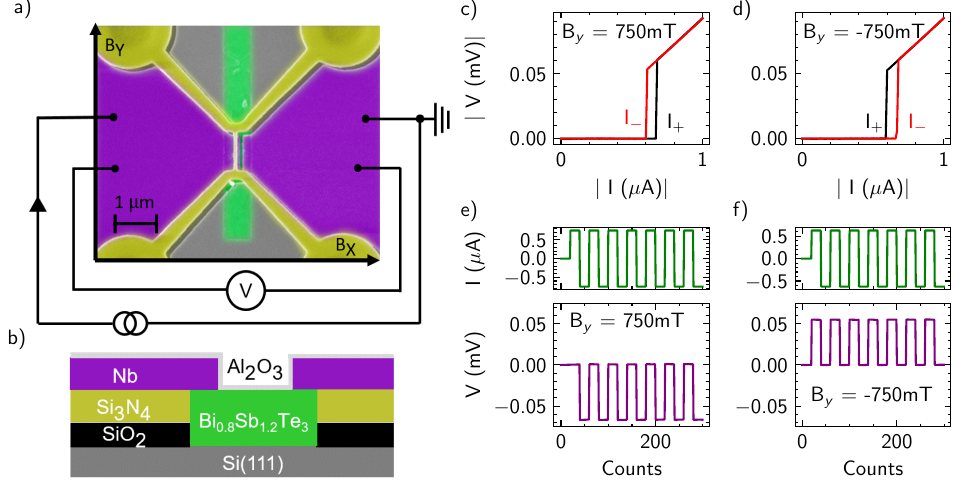}
     \caption{Sample layout and measurement of the Josephson diode effect of junction JJ$_1$. \textbf{a)} Scanning electron micrograph including the measurement circuit. \textbf{b)} Layer stack corresponding to a line cut through the junction along the transport direction. \textbf{c)} $IV$-characteristic at an applied magnetic field  $B_y$ of \SI{750}{mT} in $y$-direction perpendicular to transport. Here, $I_+$ exceeds $|I_-|$. \textbf{d)} Corresponding plot at $B_y=$\SI{-750}{mT}, with now $|I_-|$ exceeding $I_+$. \textbf{e)} Oscillating current of $\pm 600\,$nA at $B_y=$\SI{750}{mT} and corresponding voltage drop response. The junction is in the superconducting state and in the resistive state for positive and negative currents, respectively. \textbf{f)} Oscillating current of $\pm 600\,$nA at $B_y=$\SI{-750}{mT} and corresponding voltage drop. The opposite behavior to e) is recorded.}
    \label{fig:Sample_Collage}
\end{figure*}

\noindent We present low-temperature measurements of the Josephson diode effect in $\mathrm{Bi_{0.8}Sb_{1.2}Te_3}$ Josephson junctions fabricated by a combination of selective-area growth and shadow mask evaporation \cite{Sch_ffelgen_2019,Jalil_2023}. This approach allows for the in-situ fabrication of Josephson junctions with very high interface transparency, important for the study of the superconducting proximity effect. We analyze the behavior of the junction as a function of magnetic field, temperature, and microwave radiation and perform a detailed analysis of the diode effect. Two probed junctions show a pronounced diode effect. The effect is  stable over multiple switching cycles and can be reversed by inverting the polarity of an in-plane magnetic field perpendicular to the current direction. Our analysis of the temperature dependence of the critical current reveals that the supercurrent is carried to a large extent by topological surface states \cite{Sch_ffelgen_2019,Schmitt2022}.  In addition, the presence of half-integer Shapiro steps indicates a non-sinusoidal current-phase relationship in the junction. Based on these observations, we propose that the origin of the diode effect lies in the proximized topological surface states.

\section{Experimental}

\begin{figure*}[tb]
    \centering
    \includegraphics[width=0.99\linewidth]{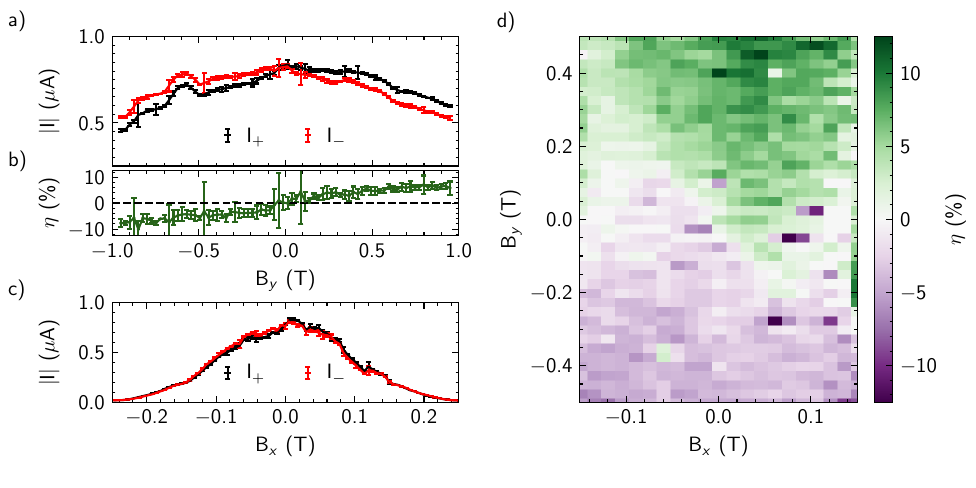}
    \caption{Full analysis of the Josephson diode effect in JJ$_{1}$ under application of magnetic fields in the $x$-$y$ plane. \textbf{a)} Critical currents $|I_-|$ and $I_+$ as a function of applied fields in the $y$-direction, i.e. perpendicular to transport. \textbf{b)} Diode rectification factor $\eta$ as a function of a magnetic field applied in $y$-direction. The errors for $\eta$ are determined using standard error propagation rules. \textbf{c)} Critical currents $|I_-|$ and $I_+$ as a function of applied fields in the $x$-direction, parallel to transport. The critical current diminishes to zero for fields larger than \SI{250}{mT}. No Josephson diode effect is present. \textbf{d)} Map of the diode rectification factor $\eta$ as a function of $B_x$ and $B_y$.}
    \label{fig:Diode_Collage}
\end{figure*}

\noindent The samples are fabricated by selective-area growth combined with shadow mask evaporation \cite{Sch_ffelgen_2019,Jalil_2023}. This approach allows the fabrication of samples with arbitrary geometries and excellent interface transparency between the topological insulator and the parent superconductor. First, a \SI{10}{nm} thick layer of SiO$_2$ followed by a \SI{25}{nm} thick layer of Si$_3$N$_4$ are deposited on a Si(111) substrate by thermal oxidation and plasma-enhanced chemical vapor deposition (PECVD), respectively to form the selective area growth mask. Trench widths of \SI{400}{nm} are prepared using electron beam lithography and reactive ion etching (RIE). A second stack of \SI{300}{nm} SiO$_2$ and \SI{100}{nm} Si$_3$N$_4$ is then deposited by PECVD to form the bridge for  shadow evaporation. The Si$_3$N$_4$ layer is patterned into the bridge shape by electron beam lithography and RIE. The sample is then etched with hydrofluoric acid to form a suspended bridge that acts as a shadow mask over the trench. The 18-nm-thick $\mathrm{Bi_{0.8}Sb_{1.2}Te_3}$ topological insulator film is grown by rotating the sample around its normal axis to ensure uniform deposition under the shadow mask. Next, \SI{50}{nm} thick Nb electrodes are deposited in situ at an angle without rotating the sample. The shadow mask provides a gap between the two Nb electrodes, effectively patterning the Josephson junction. The junction is capped with a \SI{5}{nm} thick layer of Al$_2$O$_3$ under rotation to prevent oxidation. Finally, the electrode shape is defined ex situ using an SF$_6$ RIE process, leaving the junction area and nanoribbon unaffected. A false-color scanning electron microscope (SEM) image is shown in Fig.~\ref{fig:Sample_Collage} a). For all measured junctions the width and length of the weak-link are \SI{1}{\upmu m} and \SI{100}{nm}, respectively. The corresponding layer stack along the $x$-axis of the junction is shown in Fig.~\ref{fig:Sample_Collage} b). The critical temperature $T_\mathrm{c}$ of the Nb film is determined to be approximately \SI{8.5}{K} resulting in a superconducting gap  $\Delta$ of \SI{1.3}{meV} \cite{Carbotte1090}. 

\noindent The sample characteristics were measured in a dilution refrigerator with a base temperature of $T \approx \SI{10}{mK}$. Figure~\ref{fig:Sample_Collage} a) also shows the measurement configuration, with the current $I$ applied between the two terminals and the voltage $V$ measured in a quasi four-point measurement scheme. The differential resistance $dV/dI$ is measured using a lock-in amplifier by adding a \SI{10}{nA} AC current to the applied DC current. A vector magnet (6-1-1\,T) is used to apply the magnetic field in all three Cartesian directions. The radio frequency is supplied by a standard radio frequency source using an antenna in close vicinity to the sample. 

In total five devices, i.e. JJ$_1$ to JJ$_5$, on three separate silicon chips were studied. All junctions are equally fabricated and from the same growth run. Two out of the five junctions, i.e. JJ$_1$ and JJ$_2$ showed the Josephson diode effect. The other three samples did not show a significant diode effect, presumably because of their smaller critical current and hence an insufficient number of transport channels \cite{Reynoso2012}. We attribute this to the selective-area growth approach, i.e. the different defect densities in the individually etched topological insulator trenches (see Supplementary Material). The parameters of all junctions are given in the Supplementary Material. In the following we discuss the Josephson diode effect of junction JJ$_1$, while the data of JJ$_2$ is given in Supplementary Material.

\section{Results}

\noindent The critical current $I_\mathrm{c}$ of junction JJ$_1$ is measured at zero magnetic field at \SI{10}{mK} and resulted in a value of \SI{0.880}{\upmu A}. The normal state resistance $R_\mathrm{N}$ of \SI{111}{\Omega} is evaluated by a linear regression of the junctions Ohmic behavior at voltage biases larger than the superconducting gap 2$\Delta$. Here, $2\Delta$ is extracted from the critical temperature of the Nb film $T_\mathrm{c} \approx 8.5\,\mathrm{K}$ using $\Delta = 1.764\,k_B\,T_c \approx 1.3\,$meV. The excess current $I_\mathrm{exc}$ of \SI{0.945}{\upmu A} is gained from the intercept of the linear regression at zero voltage. The transparency $\tau$, as a figure of merit to evaluate the interface quality between superconductor and the weak-link material in a Josephson junction, is obtained by fitting the analytical calculation according to the work of Niebler et al.~\cite{Niebler2009}, which is based on the Octavio-Tinkham-Blonder-Klapwijk model \cite{Octavio_1983,Flensberg_1988}. Junction JJ$_1$ exhibits a transparency of $\tau =$ \SI{68.61}{\%}, which is comparable to previous similarly fabricated samples \cite{Behner2024,Rosenbach_2021,Schmitt2022,Jalil2024}. 

\noindent First, we show that the junction JJ$_1$ indeed behaves as a Josephson diode. The performance of the Josephson diode is quantified by its rectification factor defined as: $\eta = \delta I_c / (I_+ + |I_-|)$, where $\delta I_c= (I_+ - |I_-|)$. To observe the Josephson diode effect, a magnetic field $B_y$ is applied along the $y$-axis, i.e. perpendicular to the current direction, (cf. Fig.~\ref{fig:Sample_Collage} a)) and the critical currents $|I_-|$ and $I_+$ for negative and positive biases are measured, respectively. Figure~\ref{fig:Sample_Collage} c) shows the current-voltage ($IV$) characteristics of the sample for an applied magnetic field of \SI{750}{mT}. The field is directed in-plane perpendicular to the current direction. For positive field $I_+$ = \SI{660}{nA} exceeds $ |I_-|$ = \SI{589}{nA} yielding a diode rectification factor of about $\eta\,\approx \SI{5.5}{\%}$. When switching the magnetic field direction the diode effect inverts. This can be seen in Fig.~\ref{fig:Sample_Collage} d), which depicts the $IV$-characteristics under application of $B_y=\,$\SI{-750}{mT}. Now, $|I_-|$ exceeds $I_+$ yielding $\eta\,\approx \SI{-5.6}{\%}$. In order to demonstrate the switching between superconductive and resistive state, the bias current is varied between $\pm$\SI{600}{nA}, i.e. a current magnitude between the two critical currents, while the voltage drop is recorded. This is shown in Figs.~\ref{fig:Sample_Collage} e) and f) for both field set points, respectively. For $B_y=\,$\SI{750}{mT}, the junctions is in the superconducting state for a positive but in the resistive state for negative bias current. At the opposite field $B_y=\,$\SI{-750}{mT}, this behavior switches, and the junction is in the superconducting state for a negative but in the resistive state for a positive bias current.

\begin{figure}[t]
    \centering
    \includegraphics[width=0.99\linewidth]{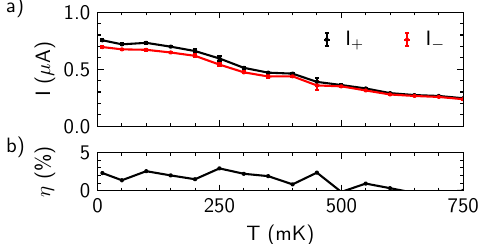}
    \caption{a) Critical currents $|I_-|$ and $I_+$ at $B_y=\,$\SI{750}{mT} as a function of temperature. b) Corresponding diode rectification factor $\eta$ as a function of temperature.}
    \label{fig:Diode_Temp}
\end{figure}

\noindent Next, we analyse the evolution of the Josephson diode effect over the whole range of magnetic fields in the $B_x$-$B_y$ plane. Since the axes of current flow, inversion symmetry breaking and time reversal symmetry breaking must all be perpendicular to each other. As the inversion symmetry is broken by the device layout in the $z$-direction, and current flow takes place in the $x$-direction we expect diode behavior only for fields applied in the $B_y$ direction. Figure~\ref{fig:Diode_Collage} a) shows the evolution of $I_-$ and $I_+$ of junction JJ$_{1}$ as a function of the applied magnetic field $B_y$ in the $y$-direction. The magnetic field is swept over a range of \SI{\pm 1}{T} with the absolute values of the critical current decreasing only by a factor of about two. This is consistent for all measured junctions and is attributed to the direction of the magnetic field with respect to the junction, i.e. the small cross section given by the thickness of the topological insulator nanoribbon and the junction length, penetrated by the magnetic field \cite{Cuevas2007,Guenel2012}. One finds that a finite diode effect develops already for fields larger than \SI{100}{mT}. The sign of the diode effect switches with inverting the magnetic field direction. As a statistical spread of the critical current was noticed at zero field, a statistical analysis of the effect is carried out. For each point in the plot, the corresponding $IV$-characteristics are measured five times each, averaged and plotted with the corresponding error bars. A full analysis of the statistical spread of the critical current at selected magnetic fields along the $x$- and $y$-axis is given in the Supplementary Material. One finds that the size of the effect is larger than the spread of the error on the critical current. In Fig.~\ref{fig:Diode_Collage} b) the diode rectification factor $\eta$ is plotted as a function of magnetic field $B_y$. The maximum diode rectification factor is determined to be $\eta_\mathrm{max} = $\SI{7}{\%}. Especially for higher fields the statistical spread shrinks and a stable diode effect is recorded. Figure~\ref{fig:Diode_Collage} c) shows the critical currents $|I_-|$ and $I_+$ of JJ$_{1}$ as a function of magnetic field applied in $x$-direction. The critical currents decrease to zero over a span of \SI{250}{mT}. As expected, no finite diode effect is present, since the magnetic field is aligned with the current direction. In the map shown in Fig.~\ref{fig:Diode_Collage} d) the diode rectification factor $\eta$ is plotted as a function of the magnetic fields in $y$- and $x$-directions. It can be clearly seen that the diode effect occurs only for magnetic fields applied along the $y$-direction, while magnetic fields in $x$-direction have basically no effect on $\eta$. In Fig.~\ref{fig:Diode_Temp}, the critical currents $|I_-|$ and $I_+$ are given as a function of temperature at an applied in-plane field of $B_y=\,$\SI{750}{mT}. 
It can be seen that the diode effect is present up to temperatures of about \SI{400}{mK}. This is consistent with observations that higher harmonic terms in the current-phase relationship disappear with increasing temperature \cite{Turini_2022}.

\begin{figure*}[t]
    \centering\includegraphics[width=0.99\linewidth]{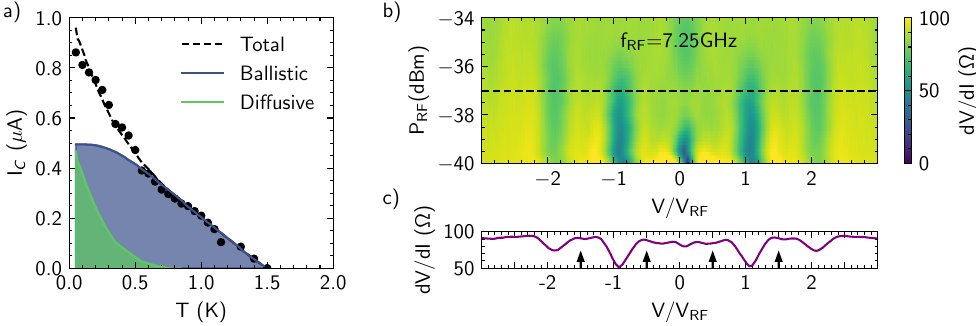}
    \caption{\textbf{a)} Temperature dependence of the critical current of junction JJ$_1$. The dashed black line shows the calculated critical current. It consists of ballistic (blue) and diffusive (green) contributions. \textbf{b)} Differential resistance $dV/dI$ as a function of the applied microwave power $P_\mathrm{RF}$ and the DC voltage normalized by the characteristic voltage $V_\mathrm{RF}=hf_\mathrm{RF}/2e$. Vertical lines in the color map correspond to dips in the differential voltage of the sample associated with plateaus in the $IV$ characteristics of the junctions. Dips are observed at integer positions of V$_\mathrm{RF}$. In addition, slightly less pronounced dips can be seen at half integer values of $V_\mathrm{RF}$. These fractional Shapiro steps are often considered as a signature of a non-sinusoidal current-phase relationship. \textbf{c)} Line cut of b) at $P_\mathrm{RF} = \SI{-37}{dBm}$. Arrows in the plot indicate the position of the fractional Shapiro steps at $n=\pm 0.5$ and $\pm 1.5$.}
    \label{fig:Shapiro}
\end{figure*}

\noindent Indeed, in order to exhibit a diode effect, the samples have to have a large number of transport channels \cite{Reynoso2012}. We therefore attribute the fact that only junctions JJ$_1$ and JJ$_2$ exhibit a sufficient number of channels in order to show the diode effect. According to the theoretical models, a non-sinusoidal current-phase relationship is required for a weak-link junction to exhibit an Josephson diode effect \cite{Reynoso2012,Yokoyama2013,Yokoyama2014,Dolcini2015,Yuan2022,Liu2024,Meyer2024}. Indeed, using an asymmetric superconducting quantum interference device, a highly skewed current-phase relationship was measured for a topological insulator weak-link, which was attributed to quasi-ballistic transport in the topological surface states \cite{Kayyalha2020}. In order to identify the contributing transport channels for our samples we measured the temperature dependence of the critical current for JJ$_1$ (see Figure~\ref{fig:Shapiro} a)). We used the clean limit Eilenberger equations to model the ballistic component of the surface states \cite{Eilenberger1968,Galaktionov2022}, while the Usadel equations were employed for the diffusive component \cite{Usadel1970}. Following the reasoning given by Sch\"uffelgen \textit{et al.} \cite{Sch_ffelgen_2019} the critical current is given by a diffusive and ballistic contribution, with the diffusive contribution diminishing at around \SI{0.7}{K}. Beyond that value, the supercurrent is only carried by the ballistic channel.

\noindent At low temperatures, weak-link junctions containing a ballistic contribution, should exhibit a non-sinusoidal current-phase relationship \cite{Golubov2004}. In order to check for this, we have recorded the $IV$-characteristics of our junctions under microwave radiation to detect Shapiro steps. For these measurements, the presence of half-integer Shapiro steps in addition to integer steps is considered as a signature of a non-sinusoidal current-phase relationship \cite{Snyder2018,Raes2020,Koelzer2023,Iorio2023}. 

\noindent Figure~\ref{fig:Shapiro} b)    
shows the differential resistance $dV/dI$ of junction JJ$_{1}$ as a function of the applied microwave power $P_\mathrm{RF}$ and the normalized measured voltage drop $V/V_\mathrm{RF}$, where $V_\mathrm{RF}=hf_\mathrm{RF}/2e$ with $f_\mathrm{RF}$ the RF frequency, $h$ the Planck constant and $e$ the electron charge. Shapiro steps in the $IV$ characteristics are expected to appear at voltages $V= n \cdot V_\mathrm{RF}$, with $n$ an integer number. These steps correspond to dips at integer values of $V/V_\mathrm{RF}$ in the differential resistance, which are indeed found in Fig.~\ref{fig:Shapiro} b) at $n=0, \pm 1$, and $\pm2$. In addition, less pronounced dips can be seen in between at $n=\pm 0.5$ and $\pm 1.5$, corresponding to half integer Shapiro steps. To emphasize their visibility, a line cut at \SI{-37}{dBm} is plotted in Fig.~\ref{fig:Shapiro} c) showing $dV/dI$ as a function of $V/V_\mathrm{RF}$. The dips at $n=\pm 0.5$ and $\pm 1.5$ are marked with vertical arrows. Together with the temperature dependent measurements, showing a ballistic critical current contribution, we can conclude that the current-phase relation is non-sinusoidal, which is an essential prerequisite for the occurrence of the Josephson diode effect.  

\noindent Taking into account the results of the temperature-dependent measurements and the measurements under microwave irradiation, we come to the following conclusions for the presence of the diode effect in our Josephson junctions. The temperature dependence of the critical current indicates the presence of a ballistic contribution to the critical current. We attribute this contribution to the existence of proximitized surface states in our junction \cite{Sch_ffelgen_2019,Schmitt2022}. The presence of  ballistic modes due to the topological surface states of our topological insulator layer is also confirmed by measurements of the Aharonov--Bohm effect on ring structures \cite{Behner2023}. It is therefore plausible to assume that our junction reside in the long ballistic junction limit since the junction length $L\approx \SI{100}{nm}$ is much longer than the coherence length of Nb $\zeta\approx\mathcal{O}(\SI{10}{nm})$ \cite{Asada1969}. This results in a non-sinusoidal, i.e. a skewed, current-phase relation \cite{Schaepers2001,Golubov2004,Murani_Thesis}. Indeed, from the appearance of half-integer Shapiro steps we can conclude that our junctions have a non-sinusoidal current-phase relation \cite{Snyder2018,Raes2020,Koelzer2023,Iorio2023}.  For topological insulator based junctions, a skewed current-phase relation has also been confirmed by an alternative measurement scheme using an asymmetric SQUID structure \cite{Kayyalha2020, Surendran2023}. However, a skewed current-phase relation is not sufficient to obtain a Josephson diode effect. In addition, an anomalous phase shift is required, which is gained by breaking time reversal and inversion symmetry through the presence of a Zeeman field and spin-orbit coupling, respectively. For junctions in the ballistic regime, including these symmetry breakings lead to an anomalous current-phase relation whose Fourier expansion includes both cosine and sinusoidal terms. The anomalous phase shift results from the finite-momentum Cooper pairing induced by the shift of the Fermi contours \cite{Yuan2022,Lotfizadeh_2024}. As a result, not only a simple phase shift is observed, but also a difference in the positive and negative current branches of the current-phase relation, i.e. a Josephson diode effect. It is well known that topological insulator materials exhibit strong spin-orbit coupling so that the conditions for a diode effect are fulfilled. Indeed, for Bi$_2$Se$_3$ Josephson junctions, an anomalous phase shift has been observed when an in-plane magnetic field is applied \cite{Chen2018,Assouline2019}. 

\section{Conclusion}

\noindent In conclusion, we have demonstrated that Nb/$\mathrm{Bi_{0.8}Sb_{1.2}Te_3}$/Nb Josephson junctions made of selectively-grown topological insulator weak links can exhibit nonreciprocal charge transport in the superconducting regime under the application of an in-plane magnetic field perpendicular to the transport. The superconducting diode effect has been studied in the whole range of in-plane magnetic fields, and these measurements confirm that the diode effect is present when the magnetic field is oriented perpendicular to the current direction. Based on temperature dependent measurements of the critical current, we conclude that the origin of the effect is the proximization of spin-orbit coupled ballistic surface states, which experience a Zeeman shift due to the applied magnetic field. With the presence of fractional Shapiro steps, the junctions also provide evidence for the existence of a non-sinusoidal current-phase relationship, necessary for the Josephson diode effect. So far, the superconducting diode effect in topological insulators has been studied only to a limited extent. However, since it seems to be the result of proximized surface states, it expands the toolbox of experiments that can be used in the ongoing search for Majorana zero-modes.


\section{Acknowledgments}

\noindent We thank H. Kertz for technical assistance, and F. Lentz and S. Trellenkamp for electron beam lithography. We are grateful for fruitful discussions with K. Moors and T. Wakamura. This work was partly funded by the Deutsche Forschungsgemeinschaft (DFG, German Research Foundation) under Germany’s Excellence Strategy - Cluster of Excellence Matter and Light for Quantum Computing (ML4Q) EXC 2004/1 – 390534769 as well as financially supported by the Bavarian Ministry of Economic Affairs, Regional Development and Energy within Bavaria’s High-Tech Agenda Project "Bausteine für das Quantencomputing auf Basis topologischer Materialien mit experimentellen und theoretischen Ansätzen" (grant no. 07 02/686 58/1/21 1/22 2/23). Furthermore, this work is supported by the QuantERA grant MAGMA and by the Deutsche Forschungsgemeinschaft (DFG, German Research Foundation) under Grant No. 491798118. 

\newpage
\putbib[bu1.bbl]  
\end{bibunit}

\clearpage
\widetext

\titleformat{\section}[hang]{\bfseries}{\MakeUppercase{Supplemental Note} \thesection:\ }{0pt}{\MakeUppercase}
\begin{bibunit}[apsrev4-1]
\setcounter{section}{0}
\setcounter{equation}{0}
\setcounter{figure}{0}
\setcounter{table}{0}
\setcounter{page}{1}
\renewcommand{\thesection}{\arabic{section}}
\renewcommand{\thesubsection}{\Alph{subsection}}
\renewcommand{\theequation}{S\arabic{equation}}
\renewcommand{\thefigure}{S\arabic{figure}}
\renewcommand{\figurename}{Supplemental Figure}
\renewcommand{\tablename}{Supplemental Table}
\renewcommand{\bibnumfmt}[1]{[S#1]}
\renewcommand{\citenumfont}[1]{S#1}

\begin{center}
\textbf{\large Supplemental Material: Superconducting Diode Effect in Selectively-Grown Topological Insulator Based
Josephson Junctions}
\end{center}

\section{Junction characteristics}

\begin{figure}[h]
    \centering
    \includegraphics[width=0.5\linewidth]{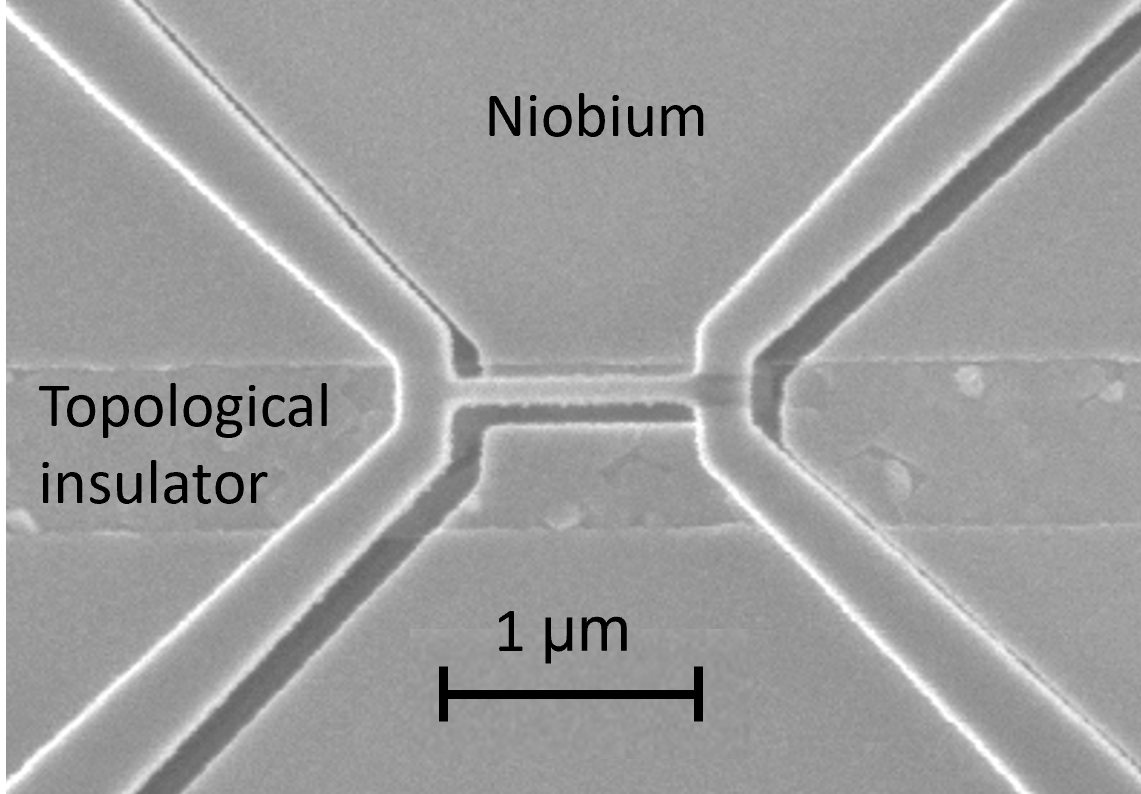}
    \caption{Scanning electron micrograph of the weak-link of Josephson junction JJ$_3$.}  
    \label{fig:Defects}
\end{figure}

A total of five devices exhibited Josephson junction behavior in the same cool-down. They will be referred to as JJ$_1$ to JJ$_5$ in the following text. The measured junction parameters are given in the Supplemental Table \ref{tab:Junction-parameters-supp}. Notice that the properties of the devices vary to a relatively large extent. We attribute this to the varying defect densities in the topological insulator weak-link of the different devices. Supplemental Figure~\ref{fig:Defects} shows a scanning electron microscope image of JJ$_3$. It shows that the material in the weak-links has a number of crystal defects, which in turn can significantly affect the properties of the weak-link.
\begin{table}[h]
    \centering
    \resizebox{0.5\linewidth}{!}{%
    \begin{tabular}{||c | c c c c c||}
         \hline
& JJ$_{1}$ & JJ$_{2}$ & JJ$_{3}$ & JJ$_{4}$ & JJ$_{5}$ \\ 
\hline\hline
$I_\mathrm{c}$ & 0.880 $\upmu$A & 0.375\,$\upmu$A & 0.120$\upmu$A & 0.069$\upmu$A & 0.048$\upmu$A \\ 
$R_\mathrm{N}$ & 111\,$\Omega$ & 113\,$\Omega$ & 418\,$\Omega$ & 1133\,$\Omega$ & 1134\,$\Omega$ \\
$I_\mathrm{exc}$  &0.945\,$\mu$A & 0.841\,$\mu$A & 0.717\,$\mu$A & 0.856\,$\mu$A &  0.871\,$\mu$A \\
$\tau$ & 68.61\,\% & 68.61\,\% & 67.05\,\% &  68.45\,\% & 68.48\,\% \\
$\eta_\mathrm{max}$ & \SI{7}{\%} & \SI{43}{\%} & \SI{0}{\%} & \SI{0}{\%} & \SI{0}{\%} \\
 \hline
    \end{tabular}}
    \caption{Parameters of Josephson junctions JJ$_1$ to JJ$_5$: $I_\mathrm{c}$ critical current, $R_\mathrm{N}$ normal state resistance, $I_\mathrm{exc}$ excess current, $\tau$ transparency, and $\eta_\mathrm{max}$ maximum diode rectification factor.}
    \label{tab:Junction-parameters-supp}
\end{table}

Two of the five devices show a finite Josephson diode effect. We attribute this to the varying critical currents in the different devices. As stated in the main text, we attribute the fact, that only two devices show a Josephson diode effect to the non-sufficient amount of transport channels in JJ$_3$ to JJ$_5$ \cite{Reynoso2012}. The transparency is relatively low compared to transparency observed in similar devices fabricated by us \cite{Behner2024}. This is the result of an increased amount of scattering at the superconductor-topological insulator interface due to the relatively large amount of crystal defects at the interface. As can be seen in Supplemental Figure ~\ref{fig:Defects}, the topological insulator material exhibits crystal defects, which will be different for every device, therefore leading to different defect densities. A different defect density can have a large effect on the properties of the single device.
\section{Additional Data Junction JJ$_1$}

\subsection{Statistics}

\begin{figure*}[h]
    \centering
    \includegraphics[width=0.99\linewidth]{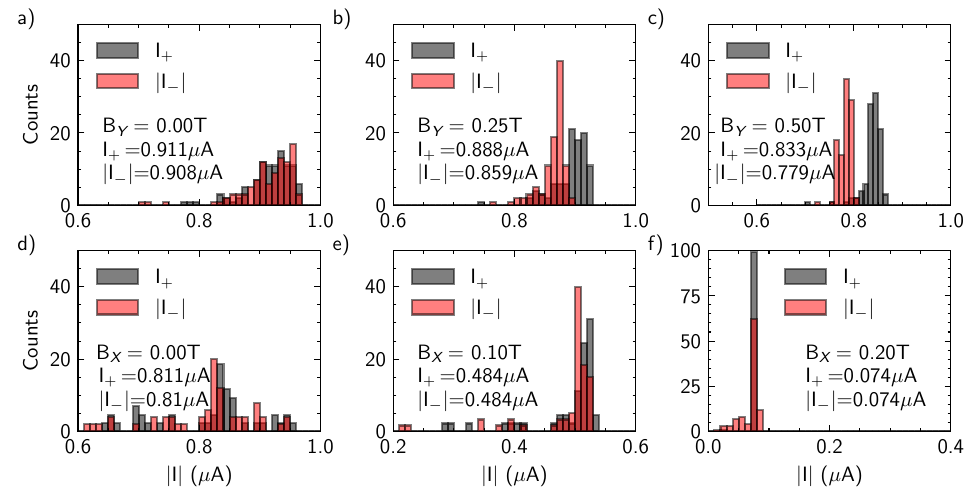}
    \caption{Statistical analysis of the switching current for JJ1. \textbf{a)} The critical currents $I_+$ and $|I_-|$ are measured 100 times respectively at the respective field value, here B$_y$=\SI{0}{T}. The result is plotted as a histogram to see the statistical variance of the current values. The spread of the measured current values is rather broad and the distributions overlaps heavily for $|I_-|$ and $I_+$. Hence, no statistically relevant diode effect is recorded. \textbf{b)} Spread of $|I_-|$ and $I_+$ at B$_y$=\SI{0.25}{T}. The two peaks for $|I_-|$ and $I_+$ separate into two independent distributions giving an rectification factor of around 1.7\%. \textbf{c) } For B$_y$=\SI{0.50}{T}, the distribution have completely separate and the statistical spread is minimized yielding an rectification factor $\eta$ = \SI{3.3}{\%}. \textbf{d)} Statistical spread of $|I_-|$ and $I_+$ at B$_x$=\SI{0.00}{T}. Note, that the difference in the distribution for B$_y$=\SI{0.00}{T} and B$_x$=\SI{0.00}{T}, which is attributed to a trapped flux, lowering the effectively the critical current. \textbf{e)} and \textbf{f)} Current distribution for B$_x$=\SI{0.1}{T} and \SI{0.2}{T}, respectively. For both fields no relevant difference in $|I_-|$ and $I_+$ are recorded.}
    \label{fig:Statistics}
\end{figure*}
In order to map out the statistical spread of the critical current of JJ$_1$, measurements at different magnetic field set-points were carried out. This is depicted in Supplemental Figure~\ref{fig:Statistics}. Here, 100 measurements of the critical currents $|I_-|$ and $I_+$ were recorded at three different set-points of B$_y$ and B$_x$, respectively, and plotted as a histogram. Supplemental Figures \ref{fig:Statistics} a) to c) depict the distribution of $|I_-|$ and $I_+$ at B$_y =\,$ \SI{0.0}{}, \SI{0.25}{}, and \SI{0.50}{T}, respectively. At zero field the spread is rather broad and the distributions for $|I_-|$ and $I_+$ overlap largely. As expected, no statistically relevant Josephson diode effect is observed. In contrast, for \SI{0.25}{}, and \SI{0.50}{T} the spread decreases and the distributions move away from one another on the current axis. Hence, yielding a statistically relevant diode effect with a rectification factor $\eta$  of \SI{1.7}{\%} and \SI{3.3}{\%}, respectively. For the set of magnetic fields $B_x$ in $x$-direction shown in Supplemental Figures~\ref{fig:Statistics} d) to f), we find that at B$_x$ = \SI{0.0}{T} the distribution is rather broad. The difference between the previous measurement at zero field shown in a) is attributed to a trapped flux, as the measurement in $x$-direction is performed after the measurement in $y$-direction. In contrast to the $y$-direction, for B$_x$ = \SI{0.1}{T} and \SI{0.2}{T},  the distributions largely overlap, as can be seen in Supplemental Figures~\ref{fig:Statistics} e) and f), respectively. Therefore, the junction does not show a statistically relevant Josephson diode effect for fields parallel to the current transport direction.

\subsection{Out of-plane magnetic field}

\begin{figure}[]
    \centering
    \includegraphics[width=0.49\linewidth]{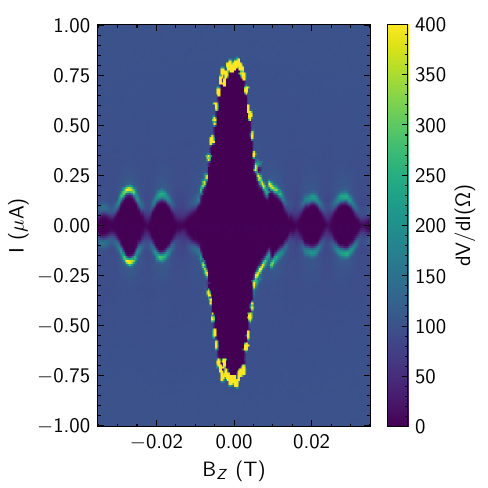}
    \caption{Fraunhofer pattern exhibited by JJ$_1$ under the application of a out-of-plane magnetic field $B_z$. The differential resistance $dV/dI$ of JJ$_{1}$, is shown as a function of applied current and magnetic field.}
    \label{fig:Fraunhofer_Analysis}
\end{figure}

\noindent We further analyse the behavior of the junction under application of an out-of-plane magnetic field $B_z$ while simultaneously applying a small in-plane field. Supplemental Figure~\ref{fig:Fraunhofer_Analysis} shows the differential resistance $dV/dI$ of JJ$_{1}$ as a function of applied current and out of-plane magnetic field. A small in-plane-magnetic field of $B_x=\,$ \SI{100}{mT} was applied in addition. The application of out of-plane field leads to the appearance of a Fraunhofer pattern of the critical current. The Fraunhofer pattern corresponds to an interference of the supercurrent in the area of the junction. It results in a periodic modulation of the supercurrent as a function of magnetic field. The periodicity  can be used to estimate the junction area. The dimensions of our junction can be extracted from the scanning electron microscopy image in Fig.~\ref{fig:Sample_Collage} a) given in the main text. The estimated junction area corresponds to a width of \SI{1}{\upmu m} and a length of \SI{100}{nm}. This leads to an expected period of \SI{21}{mT}. The experimentally observed periodicity is \SI{19}{mT}. The deviations of the measured and expected period is attributed to flux focusing effects \cite{Rosenbach_2021}. The additionally applied field of \SI{100}{mT} did not lead to a tilt in the Fraunhofer pattern due to the diminishing diode effect in this magnetic field range, which is in agreement with the results shown in Fig.~\ref{fig:Diode_Collage} a). 

\subsection{Additional Shapiro step measurements \label{SuppSect-Shapiro-JJ1}}

\begin{figure*}[h]
    \centering
    \includegraphics[width=0.99\linewidth]{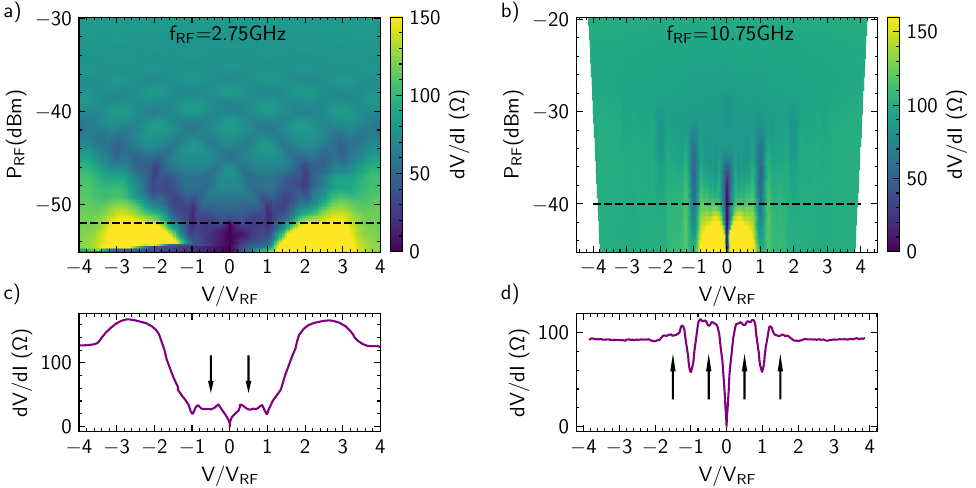}
    \caption{Additional measurements of the response of JJ$_1$ to the application of microwave radiation. \textbf{a)} Differential resistance $dV/dI$ of JJ$_1$ as a function of normalized DC-voltage $V/V_{\mathrm{RF}}$ and applied microwave power $P_\mathrm{RF}$ at a frequency of \SI{2.75}{GHz}. 
     \textbf{b)} Differential resistance $dV/dI$ of JJ$_1$ at a frequency of \SI{10.75}{GHz}. \textbf{c)} Linecut of a) at \SI{-52}{dBm}. Even though the steps are significantly smeared out, the dip indicated by the arrow suggests a fractional Shapiro step at $n = \pm 0.5$. \textbf{d)} Linecut of b) at \SI{-40}{dBm}. The arrows indicate the position of fractional steps with values $n= \pm 0.5$ and $ \pm 1.5$. Small dips in the differential resistance are visible at the suggested positions.}
    \label{fig:Shapiro_JJ1_Supp}
\end{figure*}
In the main text, only a selected set of measurements from the measurements under application microwave radiation where shown. Here, we present the evolution of the differential resistance $V/V_{\mathrm{RF}}$ of JJ$_1$ for additional frequencies. The response of JJ$_1$ to microwave radiation of frequency $f_{\mathrm{RF}}$ = \SI{2.75}{GHz} and \SI{10.75}{GHz} is depicted in Supplemental Figure~\ref{fig:Shapiro_JJ1_Supp} a) and b), respectively, which shows $dV/dI$ as a function of $V/V_{\mathrm{RF}}$ and power of the microwave radiation $P_\mathrm{RF}$. The response for both frequencies differs quite a bit. Following the argumentation of Larson \textit{et al.} \cite{Larson2020}, the different shapes can be understood in the context of the dimensionless parameter $\Omega = f_\mathrm{RF}/f_p$. Here, $f_\mathrm{p} = (1/2\pi) \sqrt{2eI_c/\hbar C}$ corresponds the bare plasma frequency of the junction, with $C$ the junction capacitance. Dependent on this parameter, the phase-evolution of the junction under application of microwave radiation will change significantly and in turn also change the Shapiro pattern. While for high $\Omega$ the plateaus are centered around fixed voltages, and their vertical range is determined by the Bessel functions, at low driving frequencies (low $\Omega$), the maps exhibit hysteretic behavior, and the plateaus corresponding to $n\neq 0$ states intersect zero bias \cite{Larson2020}.

In order to identify possible contributions of fractional Shapiro states, we analyse certain line-cuts in the color maps. Supplemental Figure \ref{fig:Shapiro_JJ1_Supp} c) shows a line cut in a) at $P_\mathrm{RF}$ = \SI{-52}{dBm}. The arrows in the graph indicates the position of the fractional step expected at $n=\pm 0.5$. At this position a dip in the differential resistance $dV/dI$ of JJ$_1$ is visible, indicating a plateau in the voltage characteristics seen as a sign of fractional Shapiro steps. The same is true for the characteristics at $P_\mathrm{RF}$ = \SI{-40}{dBm} for $f_{\mathrm{RF}}$ = \SI{10.75}{GHz} shown in Supplemental Figure \ref{fig:Shapiro_JJ1_Supp} d). Small dips in the differential resistance can be seen at the positions $n = \pm 0.5$ and  $\pm 1.5$ indicated by arrows in the graph.

\newpage


\section{Measurement Data and Analysis of Junction JJ$_2$}

\subsection{Temperature dependence of critical current}

\begin{figure*}[tb]
    \centering
    \includegraphics[width=0.99\linewidth]{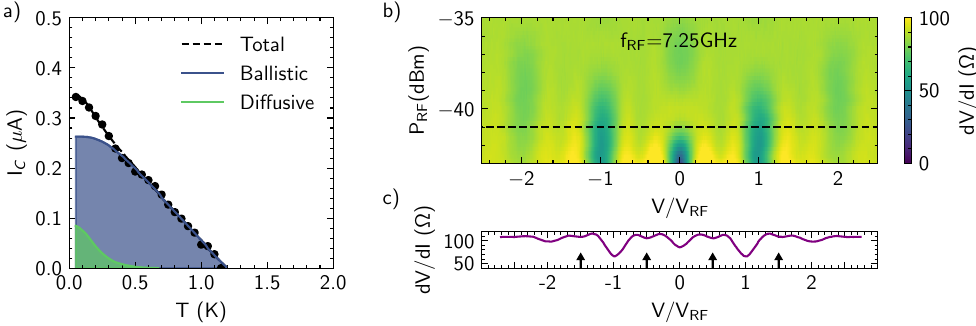}
    \caption{Temperature dependence of critical current and Shapiro step measurements of JJ$_2$. \textbf{a)} Temperature dependence of the critical current. The dashed black line shows the calculated critical current. It consists of the ballistic (blue) and diffusive (green) contributions. \textbf{b)} Differential resistance $dV/dI$ as a function of the applied microwave power $P_\mathrm{RF}$ and the DC voltage normalized by the characteristic voltage of the applied frequency $V_\mathrm{RF}=hf_\mathrm{RF}/2e$. Vertical lines in the color map correspond to dips in the differential voltage of the sample associated with plateaus in the $IV$ characteristics of the junction. Dips are observed at integer positions of $V_\mathrm{RF}$. In addition, slightly less pronounced dips can be seen at half integer values of $V_\mathrm{RF}$. \textbf{c)} Line cut of b) at $P_\mathrm{RF} = \SI{-41}{dBm}$. Arrows in the plot indicate the position of the fractional Shapiro steps at $n=\pm 0.5$ and $\pm 1.5$.}
    \label{fig:Shapiro_Supp}
\end{figure*}

Supplemental Figure~\ref{fig:Shapiro_Supp} a) shows the temperature dependence of the critical current of junction JJ$_2$. For temperatures up to \SI{0.35}{K} the curve behaves exponentially. Beyond that the critical current seems to decrease linearly. It has been shown, that this behavior is associated with the coexistence of induced superconductivity in both diffusive, as well as ballistic states in topological insulator materials \cite{Sch_ffelgen_2019}. In order to assess the contributions of both channels to the supercurrent, a fit including the ballistic as well as the diffusive contributions is performed. For the ballistic case the analysis follows the clean limit Eilenberger equations using the model presented by Galaktionov and Zaikin \cite{Eilenberger1968,Galaktionov2022}, while for the diffusive contribution the Usadel equations are used \cite{Usadel1970}. Details on the calculations of the critical current are presented in Ref.~\cite{Sch_ffelgen_2019}.

\subsection{Shapiro step measurements}

In Supplemental Figure  \ref{fig:Shapiro_Supp} b) the differential resistance $dV/dI$ of junction JJ$_{2}$ is shown as a function of the applied microwave power $P_\mathrm{RF}$ and the normalized measured voltage drop $V/V_\mathrm{RF}$, with $V_\mathrm{RF}=hf_\mathrm{RF}/2e$. Shapiro steps in the a $IV$ characteristics, corresponding to dips at integer values $n$ of $V/V_\mathrm{RF}$ in the differential resistance, are observed at with $n=\pm 1$ and $\pm 2$. In addition, less pronounced dips are found in between at $n=\pm 0.5$ and $\pm 1.5$, corresponding to half integer Shapiro steps. In the line cut shown in Supplementary Figure~\ref{fig:Shapiro_Supp} c) taken at \SI{-41}{dBm} the signatures of half-integer steps are clearly resolved. Here, the dips at $n=\pm 0.5$ and $\pm 1.5$ are marked with vertical arrows. From the presence of half integer steps we can conclude for junction JJ$_2$ that the current-phase relation is non-sinusoidal.  
\begin{figure*}[tb]
    \centering
    \includegraphics[width=0.96\linewidth]{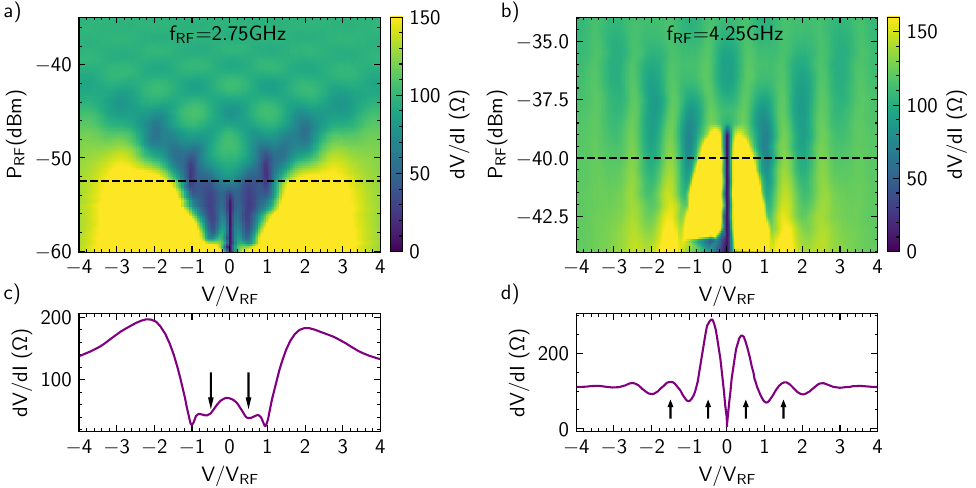}
    \caption{Additional measurements of the response of JJ$_2$ to the application of microwave radiation. \textbf{a)} Differential resistance $dV/dI$ of JJ$_2$ as a function of normalized DC-Voltage $V/V_{\mathrm{RF}}$ and applied microwave power $P_{\mathrm{RF}}$ at a frequency of \SI{2.75}{GHz}. \textbf{b)} Linecut of a) at \SI{-51.5}{dBm}, where a clear dip can be seen at a position of $V/V_{\mathrm{RF}}$ = 0.5, giving a clear indication of fractional Shapiro steps. The position of $n= \pm 0.5$ is indicated by arrows. \textbf{c)} Differential resistance $dV/dI$ of JJ$_2$ as a function of normalized DC-voltage $V/V_{\mathrm{RF}}$ and applied microwave power P$_{\mathrm{RF}}$ at a frequency of \SI{4.25}{GHz}. \textbf{d)} Line cut of c) at \SI{-40}{dBm}. Here, no indication of a contribution of fractional Shapiro steps are seen at the indicated positions of $n= \pm 0.5$ and $\pm 1.5$. }
    \label{fig:Shapiro_JJ2_Supp}
\end{figure*}

We perform the same analysis of the Shapiro response at different microwave frequencies for JJ$_2$ as for JJ$_1$ given in Supplemental Note \ref{SuppSect-Shapiro-JJ1}. The difference in the pattern also holds true for JJ$_2$ where the behavior for low and high values of $\Omega$ varies strongly. Supplemental Figure~\ref{fig:Shapiro_JJ2_Supp} a) shows the response at $f_\mathrm{RF}$ = \SI{2.75}{GHz} whereas b) gives the response at $f_\mathrm{RF}$ = \SI{4.25}{GHz}. In c) a linecut of b) is presented at $P_\mathrm{RF}$ = \SI{-51.5}{dBm}. Clear dips at $n=\pm0.5$ indicated by arrows in the graph can be seen. In contrast, in Supplemental Figure~\ref{fig:Shapiro_JJ2_Supp} d), a linecut of b) at $P_\mathrm{RF}$ = \SI{-40}{dBm}, no signatures of fractional Shapiro steps are observed.

\subsection{In-plane magnetic fields}

\begin{figure*}[h]
    \centering
    \includegraphics[width=0.99\linewidth]{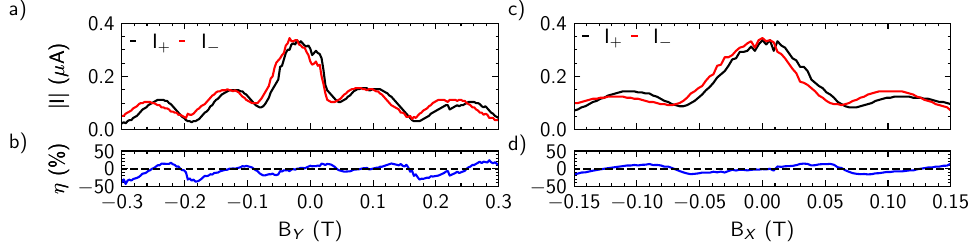}
    \caption{Analysis of the Josephson diode effect in JJ$_2$ under application of in-plane fields. \textbf{a)} Evolution of the negative and positive critical currents, i.e. $|I_-|$ and $I_+$, under application of an in-plane magnetic field B$_y$. A periodic oscillation of the critical currents can be seen, in resemblance of a Fraunhofer pattern. This indicates a misalignment of the sample with respect to the $z$-direction. A clear difference between $|I_-|$ and $I_+$ can be seen, that periodically reverses its sign. \textbf{b)} Diode rectification factor $\eta$ as a function of an applied field $B_y$. A significant diode effect with rectification factors $\eta$ up to \SI{43}{\%} are visible. The sign-reversal seems to occur with twice the frequency of what seems to resemble the Fraunhofer-pattern due to the misalignment. \textbf{c)} Evolution of $|I_-|$ and $I_+$ under application of an in-plane magnetic field $B_x$, parallel to the current. Similarly to a), a periodic oscillation of the critical currents is observed. A Josephson diode effect is observed, even though no effect is expected for field parallel to transport. This indicates a misalignment of the sample in-plane. \textbf{d)} Diode rectification factor $\eta$ as a function of a magnetic field $B_x$.} 
    \label{fig:DiodeCollageSupp}
\end{figure*}
We now perform the analysis of the diode effect in junction JJ$_2$ according to the procedures explained in the main text. Unfortunately, the sample was subject to a misalignment during the wire-bonding process, i.e. the sample was rotated in the $x$-$y$ plane by about $10$ to $15^\circ$. In addition, gluing the sample into the chip carrier also resulted in a misalignment with respect to the $z$-axis. However, despite the misalignment the sample shows a strong Josephson diode effect, even though the misalignment makes analyzing the effect in the sample to the fullest difficult.

We start by analyzing the dependence of the negative and positive critical currents, i.e. $|I_-|$ and $I_+$,  on an application of in plane fields $B_x$ and $B_y$ in the  $x$- and $y$-directions, respectively. Supplemental Figure~\ref{fig:DiodeCollageSupp} a) shows the evolution of $|I_-|$ and $I_+$ under application of an in-plane magnetic field $B_y$. Here, as shown in Fig.~\ref{fig:Sample_Collage} a) in the main text, the $B_y$-direction corresponds to the direction perpendicular to transport, whereas $B_x$ is oriented parallel to the transport direction. For both $|I_-|$ and $I_+$ a modulation of the critical current can be seen, that resembles the shape of a Fraunhofer pattern. The periodicity of the oscillation is about \SI{122}{mT} and will be used in the next subsection to determine the misalignment in the $z$-direction when comparing to the actual Fraunhofer pattern of the junction. A clear difference between $|I_-|$ and $I_+$ can be seen in the measurement. Hence, a Josephson diode effect is observed. This is shown in Supplemental Figure~\ref{fig:DiodeCollageSupp} b), where $\eta$ is plotted as a function of B$_y$. It can be seen, that a finite diode effect establishes even for a few mT applied. The rectification factor $\eta$ not only increases with increasing fields but is also subject to periodic sign changes. The periodicity seem to be about half of what we attribute to be Fraunhofer oscillations. A more detailed analysis of this sign-reversal phenomena will follow in the next section.

Supplemental Figure \ref{fig:DiodeCollageSupp} c) shows the evolution of $|I_-|$ and $I_+$ under application of an in-plane magnetic field $B_x$ in the $x$-direction. The behavior of the junction is similar to what has been analysed for fields in the $y$-direction. The junction shows flux periodic oscillations in the critical current as well as a diode effect with periodic sign-reversal (see Supplemental Figure \ref{fig:DiodeCollageSupp} d)). The existence of a diode effect for a field $B_x$ along the $x$-direction indicates that due to the above mentioned misalignment in the $x$-$y$ plane the field is not perfectly aligned along the current direction in the junction, i.e. there is a field component perpendicular to the current direction.

\subsection{Out of-plane magnetic fields}

\begin{figure*}[tb]
    \centering
    \includegraphics[width=0.99\linewidth]{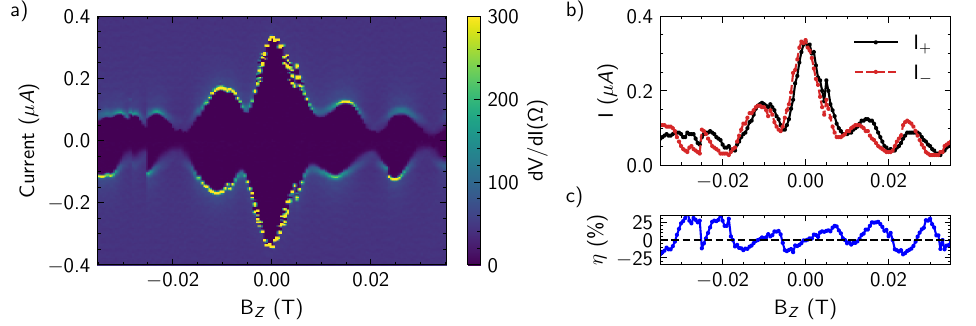}
    \caption{Analysis of the diode effect in JJ$_2$ under application of an out-of-plane magnetic field. No external in-plane field is applied. The in-plane field is a result of a misalignment. \textbf{a)} Differential resistance $dV/dI$ of JJ$_2$ plotted as a function of applied out-of-plane magnetic field and applied current. The oscillating behavior of the superconducting state is associated with the Fraunhofer-pattern of the junction. The periodicity of the oscillation amounts to about \SI{13}{mT}. A slight tilt in the patter with regards to negative and positive currents can be seen. This is due to the diode effect. \textbf{b)} The positive and negative critical currents $|I_-|$ and $I_+$ extracted from each run in a) plotted as a function of magnetic field. A clear difference between $|I_-|$ and $I_+$ can be seen as a result of the diode effect. This difference periodically reverses with applied out of plane magnetic field. \textbf{c)} The diode rectification factor $\eta$ as a function of applied out of plane magnetic field. The sign of $\eta$ changes periodically with magnetic field.}
    \label{fig:Fraunhofer_Analysis_JJ2}
\end{figure*}

Finally, we analyze the effect of an out-of plane magnetic field on the behavior with respect to the evolution of the critical current as well as the Josephson diode effect. Supplemental Figure~\ref{fig:Fraunhofer_Analysis_JJ2} a) depicts the Fraunhofer pattern of JJ$_{2}$. Here, the differential resistance $dV/dI$ of JJ$_{2}$ is plotted as a function of applied current and out-of-plane magnetic field, i.e. along the $z$-axis. Extracting the periodicity of the pattern yields a period of about \SI{13}{mT}. When extracting the area of the junction from the scanning electron microscopy image shown in Fig.~\ref{fig:Sample_Collage} a) in the main text, the period can be determined from the size of the junction. This yields an expected period of \SI{20}{mT}, with the junctions width being around \SI{1}{\upmu m} and the length being \SI{100}{nm}. As mentioned before, the deviations of the measured and expected period can be explained by flux focusing effects \cite{Rosenbach_2021}. When comparing the behavior of the junction under out-of plane fields, one notices the striking resemblance to the patterns observed under application of in-plane fields. This leads to the conclusion, that the devices is misaligned with respect to the $z$-axis. Comparing the periodicities of the modulations of $|I_-|$ and $I_+$ for $B_z$, $B_y$, and $B_x$ gives misalignment of the $x$-axis of \SI{6}{^\circ} and of the $y$-axis of \SI{5.3}{^\circ} with respect to the $z$-axis. The finite in-plane field component, due to the tilt of the sample leads to a finite diode effect while measuring the Fraunhofer pattern. This results in a tilt in the pattern, which can nicely be seen in Supplemental Figure~\ref{fig:Fraunhofer_Analysis_JJ2} a). When looking at the evolution of the critical currents $|I_-|$ and $I_+$ as a function of $B_z$ in Fig.~\ref{fig:Fraunhofer_Analysis_JJ2} b), we observe again the sign-reversal of the diode rectification factor $\eta$. This makes it evident, that the sign-reversal is an effect mediated by an out-of-plane magnetic field. A sign reversal of the Josephson diode effect has so far only been observed under application of an increased in-plane field component. Here, the sign-reversal is usually connected to anomalous $\Phi_0$-shifts and $0$-$\pi$ transitions, vortex dynamics or spin-orbit coupling induced asymmetry of the critical current under inversion of magnetic-field \cite{Lotfizadeh_2024,Margineda2023}. In the supplementary section of Ref.~\cite{Baumgartner_2022}, the effect of an out-of-plane magnetic field on the diode effect is mentioned. Here, due to the effect of an out-of-plane field on the higher harmonics of the current-phase relation  (considering only second order terms), a $\Phi_0$/2 periodic suppression of the the diode effect is expected. This fits very well to the zero-crossings that we observe in our periodic sign-reversal of the diode rectification factor $\eta$. The sign-reversal could therefore be the result of a finite phase-shift in the current-phase relation due to an out-of plane field, but would need to be studied more extensively to gain an unambiguous conclusion.

\newpage\null\newpage
\putbib[bu2.bbl] 
\end{bibunit}
\end{document}